\input phyzzx
\hoffset=0cm
\hsize=16cm
\vsize=24cm
\tenpoint
\tenrm
\sequentialequations
\singlespace

\title{\bf 
Dual Higgs Mechanism and Nonperturbative QCD 
} 

\vskip10pt

\centerline{H.~Suganuma, S.~Sasaki, H.~Ichie, H.~Toki 
and F.~Araki}

\vskip10pt

\centerline{\it 
Research Center for Nuclear Physics (RCNP), Osaka University, 
Ibaraki, Osaka 567, Japan
}

\vskip10pt

\baselineskip=12pt
\noindent
We study the dual Higgs mechanism for the nonperturbative QCD. 
We point out two kind of the ``see-saw" relations between 
electric and magnetic sectors. 
Owing to the Dirac condition $eg=4\pi$, 
the dual Ginzburg-Landau theory has the asymptotic freedom 
nature on the gauge coupling constant $e$, where 
the ``walking coupling constant" is predicted for $e$ in 
the infrared region. 
We study also QCD-monopoles and instantons, which are 
two relevant topological objects in QCD. 
Strong correlations between the QCD-monopole trajectory and 
the topological charge are found both in the continuum theory 
and in the lattice gauge theory.

\singlespace
\baselineskip=15.4pt

\vskip10pt

\noindent
{\bf 1. Introduction}

\vskip5pt

Quantum chromodynamics (QCD) is the fundamental theory 
of the strong interaction 
\REF\greiner{
W.~Greiner and A.~Sch\"afer, ``Quantum Chromodynamics", 
(Springer-Verlag, 1994).
}
\REF\itzykson{
C.~Itzykson and J.-B.~Zuber, ``Quantum Field Theory", 
(McGraw-Hill, 1985).
}
\REF\cheng{
T.~P.~Cheng and L.~F.~Li, 
``Gauge Theory of Elementary Particle Physics"
(Oxford, 1984).
}
[\greiner-\cheng].
In spite of the simple form of the QCD lagrangian,
$$
{\cal L}_{\rm QCD}=-{1 \over 2} {\rm tr} G_{\mu \nu }G^{\mu \nu }
+\bar q (\not D-m_q) q, 
\eqn\qcdlag
$$
it miraculously provides quite various phenomena 
like color confinement, dynamical chiral-symmetry breaking, 
non-trivial topologies, quantum anomalies and so on (Fig.1).
It would be interesting to compare QCD with the history of 
the Universe, because a quite simple `Big Bang' 
also created various things including galaxies, stars, lives and 
thinking reeds. 
Therefore, QCD can be regarded as an interesting 
miniature of the history of the Universe. 
This is the most attractive point of the QCD physics.

Since it is quite difficult to understand the various QCD phenomena and 
their underlying mechanism at the same time, many methods 
and models have been proposed to understand each phenomenon (Table 1).
We show in Fig.2 a brief sketch on the history of QCD and 
typical QCD effective models [\greiner]. 
In '80s, chiral symmetry breaking was the central issue. 
The chiral bag model, the NJL model and the $\sigma$ model were 
formulated with referring chiral symmetry.
In '90s, on the other hand, the confinement physics is providing 
an important current of the hadron physics, since 
recent lattice QCD studies 
\REF\hioki{
A.~S.~Kronfeld, G.~Schierholz and U.~-J.~Wiese, 
Nucl.~Phys.~{\bf B293} (1987) 461.
\nextline
S.~Hioki et al., Phys.~Lett.~{\bf B272} (1991) 326. 
\nextline
S.~Kitahara, Y.~Matsubara and T.~Suzuki, 
Prog.~Theor.~Phys.~{\bf 93} (1995) 1.
}
[\hioki] 
shed a light on the confinement mechanism: 
the origin of color confinement can be recognized 
as the dual Higgs mechanism by monopole condensation, 
and the nonperturbative QCD vacuum is regarded as 
the dual superconductor 
\REF\giacomo{
A. Di Giacomo, Nucl.~Phys.~{\bf B} (Proc.~Suppl.) {\bf 47} (1996) 136.
}
[\giacomo]. 
The dual Ginzburg-Landau theory 
\REF\suzuki{
T.~Suzuki, Prog.~Theor.~Phys. {\bf 80} (1988) 929 ; {\bf 81} (1989) 752.
}
\REF\suganumaA{
H.~Suganuma, S.~Sasaki and H.~Toki, Nucl.~Phys.~{\bf B435} (1995) 207. 
\nextline
H.~Suganuma, S.~Sasaki, H.~Toki and H.~Ichie, 
Prog.~Theor.~Phys.~(Suppl.)~{\bf 120} (1995) 57.  
}
\REF\suganumaB{
H.~Suganuma, S.~Sasaki and H.~Toki, 
Proc. of Int.~Conf. on 
``Quark Confinement and Hadron Spectrum", Como Italy, June 1994,
(World Scientific, 1995) 238.
}
\REF\suganumaC{
H.~Suganuma, H.~Ichie, S.~Sasaki and H.~Toki, 
``Confinement '95", (World Scientific, 1995) 65.
}
\REF\ichieA{
H.~Ichie, H.~Suganuma and H.~Toki, 
Phys.~Rev.~{\bf D52} (1995) 2944.
}
\REF\suganumaD{
H.~Suganuma, S.~Umisedo, S.~Sasaki, H.~Toki and O.~Miyamura,
Proc. of ``Quarks, Hadrons and Nuclei", 
Adelaide Australia, Nov. 1995, to appear in Aust. J. Phys.
}
\REF\ichieB{
H.~Ichie, H.~Suganuma and H.~Toki, Phys.~Rev. {\bf D} in press.
\nextline
H.~Ichie, H.~Monden, H.~Suganuma and H.~Toki, 
Proc. of ``Nuclear Reaction Dynamics of Nucleon-Hadron Many Body 
Dynamics", Osaka, Dec. 1995, in press.
}
[\suzuki-\ichieB] 
was formulated from QCD with this picture, and provides a useful 
framework for the study of the nonperturbative QCD.

\vskip10pt

\noindent
{\bf 2. Color Confinement and Dual Higgs Mechanism}

\vskip5pt

We briefly show the modern current of the confinement physics. 
The QCD vacuum can be regarded as the dual version of the 
superconductor, which was firstly pointed out by Nambu, 't~Hooft 
and Mandelstam 
\REF\nambu{
Y.~Nambu, Phys.~Rev.~{\bf D10} (1974) 4262.
\nextline
G.~'t~Hooft, ``High Energy Physics" 
(Editorice Compositori, Bologna 1975).
\nextline
S.~Mandelstam, Phys.~Rep.~{\bf C23} (1976) 245.
}
[\nambu].
Here, the ``dual version'' means the interchange between 
the electric and magnetic sectors.
With referring Table 2 and Fig.3, we compare the ordinary 
electromagnetic system, the superconductor and 
the nonperturbative QCD vacuum regarded as the dual superconductor. 

In the ordinary electromagnetism in the Coulomb phase, 
both electric flux and magnetic flux 
are conserved, respectively. The electric-flux conservation 
is guarantied by the ordinary gauge symmetry. 
On the other hand, the magnetic-flux conservation 
is originated from the dual gauge symmetry [\suganumaA-\ichieB], 
which is the generalized version of the Bianchi identity. 
As for the inter-charge potential in the Coulomb phase, 
both electric and magnetic potentials are Coulomb-type. 

The superconductor in the Higgs phase is characterized by 
electric-charge condensation, which leads to the Higgs mechanism or 
spontaneous breaking of the ordinary gauge symmetry, 
and therefore the electric flux is no more conserved. 
In such a system obeying the London equation, 
the electric inter-charge potential becomes short-range Yukawa-type 
similarly in the electro-weak unified theory. 
On the other hand, the dual gauge symmetry is not broken, so that the 
magnetic flux is conserved, but is squeezed like a one-dimensional 
flux tube due to the Meissner effect.
As the result, the magnetic inter-charge potential becomes 
linearly rising like a condenser.

The nonperturbative QCD vacuum regarded as the dual Higgs phase 
is characterized by color-magnetic monopole condensation, 
which leads to the spontaneous breaking of the dual gauge symmetry. 
Therefore, color-magnetic flux is not conserved, 
and the magnetic inter-change potential becomes short-range Yukawa-type.
Note that the ordinary gauge symmetry is not broken by such monopole 
condensation. Therefore, color-electric flux is conserved, 
but is squeezed like a one-dimensional flux-tube or a string 
as a result of the dual Meissner effect.
Thus, the hadron flux-tube is formed in the monopole-condensed 
QCD vacuum, and the electric inter-charge potential 
becomes linearly rising, which confines the color-electric charges 
[\suganumaA-\suganumaD].

As a remarkable fact in the duality physics, 
these are two ``see-saw relations'' between the electric and 
magnetic sectors. 
\nextline
(1) There appears the Dirac condition $eg=4\pi$ [\suganumaA] in QCD. 
Here, unit electric charge $e$ is the gauge coupling constant, 
and unit magnetic charge $g$ is the dual gauge coupling constant. 
Therefore, a strong-coupling system in one sector corresponds to 
a weak-coupling system in the other sector.
\nextline
(2) As shown in Fig.3, the long-range confinement system in one sector 
corresponds to a short-range (Yukawa-type) interaction system in the 
other sector.

Let us consider usefulness of the latter ``see-saw relation''.
One faces highly non-local properties among the color-electric 
charges in the QCD vacuum because of the long-range linear 
confinement potential.
Then, the direct formulation among the electric-charged variables 
would be difficult due to the non-locality, 
which seems to be a destiny in the long-distance confinement physics. 
However, one finds a short-range Yukawa potential 
in the magnetic sector, and therefore the electric-confinement 
system can be approximated by a local formulation 
among magnetic-charged variables. 
Thus, the confinement system, which seems highly non-local,  
can be described by a short-range interaction theory using 
the dual variables. This is the most attractive point 
in the dual Higgs theory.

Color-magnetic monopole condensation is necessary for 
color confinement in the dual Higgs theory. 
As for the appearance of color-magnetic monopoles in QCD, 't~Hooft 
proposed an interesting idea of the abelian gauge fixing
\REF\thooft{
G.~'t~Hooft, Nucl.~Phys.~{\bf B190} (1981) 455.
}
\REF\ezawa{
Z.~F.~Ezawa and A.~Iwazaki, Phys.~Rev.~{\bf D25} (1982) 2681; 
{\bf D26} (1982) 631.
}
[\thooft,\ezawa], which is defined by the diagonalization 
of a gauge-dependent variable. 
In this gauge, QCD is reduced into an abelian gauge theory with 
the color-magnetic monopole [\suganumaA,\thooft], 
which will be called as QCD-monopoles hereafter. 
Similar to the 't~Hooft-Polyakov monopole 
[\cheng] in the Grand Unified theory 
(GUT), the QCD-monopole appears from a hedgehog configuration 
corresponding to the non-trivial homotopy group 
$\pi_2({\rm SU}(N_c)/{\rm U}(1)^{N_c-1})=Z_\infty ^{N_c-1}$ 
on the nonabelian manifold. 

Many recent studies based on the lattice QCD 
with 't~Hooft abelian gauge fixing 
\REF\miyamura{
O.~Miyamura, Nucl.~Phys.~{\bf B}(Proc.~Suppl.){\bf 42} (1995) 538.
\nextline
O.~Miyamura and S.~Origuchi, ``Color Confinement and Hadrons'', 
(World Scientific, 1995) 235.
}
\REF\suganumaE{
H.~Suganuma, A.~Tanaka, S.~Sasaki and O.~Miyamura, 
Proc. of ``Lattice Field Theories '95", 
Nucl.~Phys.~{\bf B}~(Proc.~Suppl.)~{\bf 47} (1996) 302.
}
\REF\suganumaF{
H.~Suganuma, K.~Itakura, H.~Toki and O.~Miyamura, 
Proc. of ``Non-perturbative Approaches to QCD", 
Trento Italy, (PNPI-report, 1995) 224. 
}
[\hioki,\suganumaD,\miyamura-\suganumaF] 
show QCD-monopole condensation in the confinement phase, 
and strongly support abelian dominance and monopole dominance 
for the nonperturbative QCD (NP-QCD), e.g., 
linear confinement potential, dynamical chiral-symmetry breaking 
(D$\chi $SB) and instantons.
Here, abelian dominance means that QCD phenomena is described only 
by abelian variables in the abelian gauge. 
Monopole dominance is more strict, and means that 
the essence of NP-QCD is described only by the singular 
monopole part of abelian variables 
[\hioki,\suganumaD,\miyamura-\suganumaF].

Figure 4 is the schematic explanation on abelian dominance 
and monopole dominance observed in the lattice QCD. 

\noindent
(a) Without gauge fixing, it is very difficult to extract 
relevant degrees of freedom in NP-QCD. 

\noindent
(b) In the abelian gauge, only U(1) gauge 
degrees of freedom including monopole is relevant for 
NP-QCD: abelian dominance. 
Here, monopole condensation is observed as the appearance of 
a very long monopole loop in the confinement phase. 
On the other hand, off-diagonal parts does not contribute to NP-QCD.

\noindent
(c) The U(1)-variable can be separated into the regular 
photon part and the singular monopole part 
\REF\dgt{
T.~DeGrand and D.~Toussaint, Phys.~Rev.~{\bf D22} (1980) 2478.
}
[\dgt]. 
The monopole part leads to NP-QCD 
(confinement, D$\chi $SB, instanton): monopole dominance.
On the other hand, the photon part is almost trivial similar to 
the QED system.

\noindent
Thus, the condensed monopole in the 't~Hooft abelian gauge is 
nothing but the relevant collective mode for NP-QCD, 
and therefore the NP-QCD vacuum can be identified as 
the dual-superconductor in a realistic sense.

\vskip10pt

\noindent
{\bf 3. Asymptotic Freedom in the Dual Ginzburg-Landau Theory}

\vskip5pt

\noindent
{\it 3-1. Dual Ginzburg-Landau Theory}

\vskip5pt

The dual Ginzburg-Landau (DGL) theory is the QCD effective theory 
based on the dual Higgs mechanism, and can be derived from 
the QCD lagrangian with considering the QCD nature: 
monopole condensation and abelian dominance [\hioki].  
The DGL lagrangian [\suganumaD,\ichieB] 
for the pure-gauge system is described with 
the dual gauge field $B_\mu$ and the QCD-monopole field $\chi$, 
$$
{\cal L}_{\rm DGL}={\rm tr} \left\{
-{1 \over 2}(\partial _\mu  B_\nu -\partial _\nu B_\mu )^2
+[\hat{D}_\mu, \chi]^{\dag}[\hat{D}^\mu, \chi]
-\lambda ( \chi^{\dag} \chi -v^2)^2 \right\},
\eqn\dgllag
$$
where $\hat{D}_\mu \equiv \hat{\partial}_\mu +i g B_\mu$ 
is the dual covariant derivative 
including the dual gauge coupling constant $g=4\pi/e$.

The dual gauge field 
$B_\mu \equiv \vec B_\mu \cdot \vec H = B_\mu^3 T^3+B_\mu^8 T^8$ 
is defined on the dual gauge manifold 
${\rm U(1)}_m^3 \times {\rm U(1)}_m^8$ [\suganumaD,\ichieB], 
which is the dual space of the maximal torus subgroup
${\rm U(1)}_e^3 \times {\rm U(1)}_e^8$ 
embedded in the original gauge group ${\rm SU(3)}_c$.
The abelian field strength tensor is written as  
$F_{\mu \nu}={}^* (\partial\wedge B)_{\mu\nu}$, 
so that the role of the electric and magnetic fields are interchanged 
in comparison with the ordinary $A_\mu$ description.

The QCD-monopole field $\chi$ is defined as 
$\chi \equiv \sqrt{2} \sum_{\alpha=1}^3 \chi_\alpha E_\alpha$ 
[\suganumaD,\ichieB] using 
$E_1 \equiv {1 \over {\sqrt{2}}}(T_6+iT_7)$,
$E_2 \equiv {1 \over {\sqrt{2}}}(T_4-iT_5)$ and 
$E_3 \equiv {1 \over {\sqrt{2}}}(T_1+iT_2)$.
Here, $\chi_\alpha$ has the magnetic charge $g\vec \alpha$ 
proportional to the root vector $\vec \alpha$. 
In the QCD-monopole condensed vacuum with $|\chi_\alpha| =v$, 
the dual gauge symmetry ${\rm U(1)}_m^3 \times {\rm U(1)}_m^8$ 
is spontaneously broken. 
Through the dual Higgs mechanism, 
the dual gauge field $B_\mu$ acquires its mass $m_B = \sqrt{3}g v $, 
whose inverse provides the radius of the hadron flux tube [\suganumaA], 
and the dual Meissner effect causes the color-electric field 
excluded from the QCD vacuum, which leads to color confinement. 
The QCD-monopole fluctuations 
$\tilde \chi_\alpha \equiv \chi_\alpha - v$ ($\alpha$=1,2,3) 
also acquire their mass $m_\chi= 2 \sqrt{\lambda} v$ in the 
QCD-monopole condensed vacuum. 
As a relevant prediction, only one QCD-monopole fluctuation 
$\tilde \chi_ \equiv \sum_{\alpha=1}^3 \tilde \chi_\alpha$ appears as a 
color-singlet scalar glueball in the confinement phase, 
although the dual gauge field $B_\mu$ and 
the other two combinations of the QCD-monopole fluctuation 
are not color-singlet and cannot be observed 
[\suganumaD,\suganumaE,\suganumaF]. 

The DGL theory reproduces confinement properties like 
the inter-quark potential and the hadron flux-tube formation. 
We studied effects of the flux-tube breaking 
by the light quark-pair creation in the DGL theory, 
and derived the infrared screened inter-quark potential 
[\suganumaA-\suganumaC], 
which is observed in the lattice QCD with dynamical quarks. 
We studied also the dynamical chiral-symmetry breaking (D$\chi$SB), 
which is also an important feature in the nonperturbative QCD, 
by solving the Schwinger-Dyson equation for the dynamical quark 
[\suganumaA-\suganumaC]. 
The quark-mass generation is brought by QCD-monopole condensation, 
which suggests the close relation between D$\chi$SB 
and color confinement.
Thus, the DGL theory provides not only the confinement properties 
but also D$\chi$SB and its related quantities like the constituent 
quark mass, the chiral condensate and the pion decay constant 
\REF\sasaki{
S.~Sasaki, H.~Suganuma and H.~Toki, Prog.~Theor.~Phys. {\bf 94} (1995) 
373.
\nextline
S.~Sasaki, H.~Suganuma and H.~Toki, Proc. of Int.~Conf.~on 
``Baryons '95", Santa Fe, Oct. 1995.
}
[\suganumaA-\suganumaD,\sasaki].

\vskip5pt

\noindent
{\it 3-2. Asymptotic Behavior in DGL Theory}

\vskip5pt

In the abelian gauge fixing, the Dirac condition $eg=4\pi $ for the 
dual gauge coupling constant $g$ is derived [\thooft,\suganumaA] 
as a geometrical constraint on the nonabelian manifold 
using the similar argument on the GUT-monopole [\cheng]. 
Since the Dirac condition is a geometrical relation, 
it is renormalization group invariant. 
As the first category of the ``see-saw relation" in Section 2, 
$g$ becomes small for the strong coupling region 
with a large coupling constant $e$ [\suganumaA,\suganumaB]. 

The DGL theory in the pure gauge is renormalizable similar to
the abelian Higgs model [\itzykson], and is not asymptotically 
free on $g$ in view of the renormalization group: 
$g(p^2)$ is increasing function of $p^2$.
Hence, asymptotic freedom is expected for the QCD gauge coupling 
constant $e$ owing to the Dirac condition: 
$e(p^2)$ defined by $e(p^2)g(p^2)=4\pi$ is 
decreasing function of $p^2$.
Thus, the DGL theory qualitatively shows asymptotic freedom 
on the QCD gauge coupling $e$ [\suganumaA,\suganumaB]. 
This asymptotic behavior in the DGL theory is consistent 
with QCD qualitatively, 
and seems a desirable feature for an effective theory of QCD.

Next, we attempt to calculate the $\beta$-function and 
the running coupling constant 
from the polarization tensor $\Pi _{\mu \nu}^{ab}(p)$ 
of the dual gauge field $B_\mu$ in the DGL theory. 
In particular, we are interested in the infrared region 
($p \lsim 1{\rm GeV}$), where the perturbative QCD calculation 
is not reliable. 

With the dimensional regularization [\itzykson,\cheng] 
by shifting the space-time dimension as $d=4-\epsilon $, 
we calculate the simplest nontrivial radiative correction 
from the QCD-monopole loop diagrams as shown in Fig.5, 
$$
\Pi _{\mu \nu}^{ab}(p)=-{1 \over 32\pi^2}\delta ^{ab}(g\mu^{-\epsilon })^2 
(p^2g_{\mu\nu}-p_\mu p_\nu){1 \over \epsilon } + O(\epsilon ^0), 
\eqn\POLa
$$
where $g$ is the bare dual-gauge coupling and 
$\mu$ the renormalization point.
In the minimum subtraction scheme [\cheng], 
the O(1/$\epsilon $) divergence is eliminated by the counter term contribution,
$$
\Pi _{\mu\nu}^{{\rm C} ab}(p)=-(Z_3-1)\delta ^{ab}(p^2 g_{\mu\nu} - p_\mu p_\nu),
\eqn\POLb
$$
where $Z_3$ is the wave-function renormalization constant [\itzykson] 
of the dual gauge field $B_\mu$, 
$$
Z_3(\mu)=1-{(g\mu^{-\epsilon })^2 \over 32 \pi^2}{1 \over \epsilon }.
\eqn\Zfactor
$$
Because of the Ward identity $Z_1=Z_2$ [\itzykson], 
the renormalized coupling constant is simply given by 
\nextline
$g(\mu) = Z_3(\mu)^{1/2} g$. 
The $\beta $-function [\itzykson,\cheng] is then expressed as 
$$
\beta  \equiv \mu {d \over d \mu}g(\mu)
=\mu {d \over d \mu}\{Z_3(\mu)^{1/2}g\}
={1 \over 32\pi^2}g(\mu)^3+O[g(\mu)^5], 
\eqn\BETAf
$$
which determines the behavior of the running coupling $g(\mu)$ as  
$$
{1 \over g^2(\mu)}={1 \over g^2(\mu_0)}
-{1 \over 32\pi^2}\ln(\mu^2/\mu_0^2)
\eqn\RUNa
$$
within the leading order.

By summation of the multi-polarization diagrams, one obtains 
the final formula for the running coupling $g(\mu)$ 
including the higher order correction, 
$$
g^2(\mu)=g^2(\mu_0)-{1 \over 32\pi^2}\ln(\mu^2/\mu_0^2).
\eqn\RUNb
$$
In the DGL theory, the QCD gauge coupling $e(\mu)$ defined 
by $e(\mu)g(\mu)=4\pi$ is simply expressed as 
$$
e^2(\mu)=e^2(\mu_0) \left\{
1+{1 \over e^2(\mu_0)}\ln(\mu/\mu_0)
\right\}^{-1}.
\eqn\RUNc
$$

We show in Fig.6 the running coupling constants, 
$g(\mu)$ and $e(\mu)$, in the DGL theory 
with the parameter set in Ref.[\suganumaC]: $m_\chi$=1.67GeV.
Similarly in QED or the abelian Higgs model, 
we have imposed the renormalization condition as 
$g(\mu=2m_\chi)=7.9$ [ $e(\mu=2m_\chi)=1.59$ ], which 
is taken to be consistent with the parameter 
$g=6.28$ ($e=2.0$) in Ref.[\suganumaC] in the infrared region 
(see Fig.6).

As a model prediction in the DGL theory, 
the gauge coupling $e(\mu)$ behaves as ``walking coupling constant", 
which means the slowly varying running coupling, 
even in the infrared region as $\mu \lsim 1 {\rm GeV}$.

\vskip10pt

\noindent
{\bf 4. Correlation between Instanton and QCD-monopole}

\vskip5pt

\noindent
{\it 4-1. Analytical Study}

\vskip5pt

The instanton is another important topological object 
in the nonabelian gauge theory; $\pi_{3}({\rm SU}(N_c))$ =$Z_\infty $ 
\REF\diakonov{
D.~I.~Diakonov and V.~Yu.~Petrov, 
Nucl.~Phys.~{\bf B245} (1984) 259.
}
[\suganumaD,\suganumaE,\suganumaF,\diakonov]. 
Recent lattice studies [\hioki,\suganumaD,\miyamura-\suganumaF] 
indicate abelian dominance for the nonperturbative quantities 
in the maximally abelian (MA) gauge and in the Polyakov gauge. 
If the system is completely described only by the abelian field, 
the instanton would lose the topological basis 
for its existence, and therefore it seems unable to 
survive in the abelian manifold. 
However, even in the abelian gauge, nonabelian components remain 
relatively large around the QCD-monopoles, which are nothing 
but the topological defects, so that instantons 
are expected to survive only around the QCD-monopole trajectories 
in the abelian-dominant system 
[\suganumaD,\suganumaE,\suganumaF]. 

We have pointed out such a close relation between 
instantons and QCD-monopoles 
in the continuum SU(2) gauge theory 
in the Polyakov-like gauge, where $A_4(x)$ is diagonalized 
[\suganumaB,\suganumaE,\suganumaF]. 
Using the 't~Hooft symbol $\bar \eta ^{a\mu \nu }$, 
the multi-instanton solution is written as 
[\suganumaD,\suganumaE,\suganumaF]
$
A^\mu (x)=i\bar \eta ^{a\mu \nu }{\tau ^a \over 2} \partial^\nu 
\ln \left(1+\sum_k {a_k^2 \over |x-x_k|^2}\right), 
$
where $x_k^\mu  \equiv ({\bf x}_k,t_k)$ and $a_k$ denote the center 
coordinate and the size of $k$-th instanton, respectively. 
Near the center of $k$-th instanton, 
$A_4(x)$ takes a hedgehog configuration, 
$
A_4(x) \simeq i {\tau ^a ({\bf x}-{\bf x}_k)^a \over |x-x_k|^2}.
$
In the Polyakov-like gauge, 
$A_4(x)$ is diagonalized by a singular gauge transformation 
[\suganumaA,\suganumaF], 
which leads to the QCD-monopole trajectory on $A_4(x)=0$: 
${\bf x} \simeq {\bf x}_k$. 
Hence, the QCD-monopole trajectory penetrates 
each instanton center along the temporal direction. 
In other words, instantons only live along the QCD-monopole trajectory. 

In the multi-instanton system 
\REF\teper{
A.~Hart and M.~Teper, Phys.~Lett.~{\bf B} (1996).
}
\REF\polikarpov{
M.~I.~Polikarpov, plenary talk at ``Lattice Field Theories '96", 
St. Louis USA, June 1996. 
}
[\suganumaB,\suganumaE,\suganumaF,\teper,\polikarpov], 
the QCD-monopole trajectory tends to be highly complicated 
and unstable against a small fluctuation of 
the location or the size of instantons [\suganumaE,\suganumaF] 
even at the classical level. 
The QCD-monopole trajectory has loops or a folded structure, 
and the topology of the trajectory is often changed due to a small 
fluctuation of instantons 
[\suganumaE,\suganumaF].

We show in Fig.7 an example of the QCD-monopole trajectory 
in the multi-instanton system, where all instantons are 
put on the $zt$-plane for simplicity.
The contour denotes the magnitude of the topological density.
Each instanton attaches the QCD-monopole trajectory. 
As a remarkable feature in the Polyakov-like gauge, 
the monopole favors the high topological density region, ``mountain''. 
Each monopole trajectory walks crossing tops of the mountain.
On the other hand, anti-monopole with the opposite color-magnetic 
charge favors the low topological density region, ``valley". 
Thus, the strong local correlation is found between the instanton and 
the QCD-monopole.

Furthermore, quantum fluctuation should make the QCD-monopole 
world line more complicated and more unstable, 
which leads to appearance of a long complicated trajectory as a result.
Thus, instantons would contribute to promote monopole condensation, 
which is signaled by a long complicated monopole loop 
as shown in Fig.8 in the lattice QCD simulation [\hioki]

\vskip10pt

\noindent
{\it 4-2. Lattice Study}

\vskip5pt

We study the correlation between instantons and QCD-monopoles 
\REF\thurner{
S.~Thurner, H.~Markum and W.~Sakuler,
``Confinement '95", (World Scientific, 1995) 77.
}
[\miyamura,\suganumaE,\suganumaF,\thurner]
in the maximally abelian (MA) gauge and in the Polyakov gauge 
using the SU(2) lattice with  $16^4$ and $\beta=2.4$. 
The SU(2) link variable can be separated into the singular monopole 
part and the regular photon part [\miyamura,\suganumaE,\suganumaF]. 
We find that instantons and anti-instantons exist only in the 
monopole sector in the abelian gauges, 
which means monopole dominance for the topological 
charge [\miyamura,\suganumaE,\suganumaF].
Monopole dominance for the U$_{\rm A}$(1) anomaly is also expected.

We study also the finite-temperature system using the 
$16^3 \times 4$ lattice with various $\beta $ around 
$\beta _c \simeq 2.3$ 
\REF\insam{
H.~Suganuma and O.~Miyamura, Proc. of INSAM Symp. '95, 
Hiroshima Univ., Oct. 1995.
}
[\insam].
Monopole dominance for the instanton 
is found in the finite-temperature confinement phase. 
Near the critical temperature $\beta _c \simeq 2.3$, 
we observe a large reduction of the number of instantons 
and anti-instantons.
In the deconfinement phase ($\beta >\beta_c$), 
instantons vanish as well as QCD-monopole condensation. 
Hence, instantons are expected to survive only around 
the condensed QCD-monopole trajectories [\insam] (see Fig.10).

Finally, we study the correlation between instanton number 
and QCD-monopole loop length.
There appears a very long QCD-monopole loop 
in each lattice gauge configuration in the confinement phase 
as shown in Fig.8. 
Hence, a very long QCD-monopole loop is regarded as a signal of 
the confinement in the lattice QCD [\hioki]. 
As shown in Fig.9, a linear correlation is found between 
the total monopole-loop length $L$ and 
$I_Q \equiv \int d^4x |{\rm tr}(G_{\mu\nu} \tilde G_{\mu\nu})|$, 
which corresponds to the total number $N_{\rm tot}$
of instantons and anti-instantons. 

From the above results, we propose the following conjecture. 
Each instanton accompanies a small monopole loop nearby 
[\teper,\thurner], whose length would be proportional to 
the instanton size. 
When $N_{\rm tot}$ is large enough, 
these monopole loops overlap, and there appears 
a very long QCD-monopole trajectory, which bonds neighboring instantons 
as shown in Fig.10. 
Such a monopole clustering leads to monopole condensation and 
color confinement. 
Thus, multi-instanton is expected to provide a source of the color 
confinement through the monopole clustering [\suganumaE,\suganumaF].

We thank Professors O.~Miyamura and D.~Diakonov 
for their useful discussion on the instanton properties. 
The Monte Carlo simulations in this paper 
have been performed on the Intel Paragon 
XP/S(56node) at the Institute for Numerical Simulations 
and Applied Mathematics of Hiroshima University.

\baselineskip=15pt

\refout

\end

The key word for the understanding of confinement 
is the ``duality'', which is recently paid attention by 
many theoretical particle physicists after Seiberg-Witten's discovery 
on the essential role of monopole condensation for the confinement 
in a supersymmetric version of QCD 
\REF\seiberg{
N.~Seiberg and E.~Witten, Nucl.~Phys.~{\bf B426} (1994) 19; 
{\bf B431} (1994) 484. 
}
[\seiberg].